\documentclass[a4paper]{jpconf}
\usepackage{graphicx}
\usepackage{amsmath}   
\usepackage{epsfig}
\begin{document}
\title{Radiation hardness test of un-doped CsI crystals and Silicon Photomultipliers for the Mu2e calorimeter}

\author{ S Baccaro$^{1}$,  A Cemmi$^{1}$, M Cordelli$^{2}$, E Diociaiuti*$^{2,3}$, R Donghia*$^{2,3}$, S~Giovannella$^{2}$, S~Loreti$^{4}$, S Miscetti$^{2}$, M Pillon$^{4}$ and I Sarra$^{2}$}

\address{ $^1$ ENEA FSN, Casaccia R.C., Rome, Italy}
\address{$^2$ Laboratori Nazionali di Frascati - INFN, Frascati, Italy}
\address{$^3$ Universit\'a degli Studi Roma Tre, Rome, Italy}
\address{$^4$ Department of Fusion and Technology for Nuclear Safety and Security - ENEA C.R., Frascati, Italy}

\ead{eleonora.diociaiuti@lnf.infn.it, raffaella.donghia@lnf.infn.it}

\begin{abstract}
The Mu2e calorimeter is composed by 1400 un-doped CsI crystals coupled to large area UV extended Silicon Photomultipliers arranged in two annular disks. This calorimeter has to provide precise information on energy, timing and position. It should also be fast enough to handle the high rate background and it must operate and survive in a high radiation environment. Simulation studies estimated that, in the hottest regions, each crystal will absorb a dose of 300 Gy and will be exposed to a neutron fluency of $6 \times 10^{11}$~n/cm$^2$ in 3 years of running. 

Test of un-doped CsI crystals irradiated up to 900 Gy and to a neutron fluency up to $9 \times 10^{11}$~n/cm$^2$ have been performed at CALLIOPE and FNG ENEA facilities in Italy. We present our study on the variation of light yield (LY) and longitudinal response uniformity (LRU) of these crystals after irradiation. The ionization dose does not modify LRU while a 20\% reduction in LY is observed at 900 Gy. Similarly, the neutron flux causes an acceptable LY deterioration ($\leq$ 15\%). 
A neutron irradiation test on different types of SIPMs (two different array models from Hamamatsu and one from FBK) have also been carried out by measuring the variation of the leakage current and the charge response to an ultraviolet led. We concluded that, in the experiment, we will need to cool down the SIPMs to 0 $^\circ$C reduce the leakage current to an acceptable level.
\end{abstract}

\section{Introduction}
The Mu2e \cite{tdr} calorimeter is composed by 1400 pure CsI crystals coupled to large area UV extended Silicon 
Photomultipliers (SiPM), arranged in two annular disks~\cite{misc}\cite{martini}. 
It has to have a good energy, timing and position resolution and 
to operate in a high radiation environment~\cite{misc}\cite{martini}. 

Simulation studies shows that in the hottest regions each crystal 
will absorb 30 krad (300~Gy) of ionization dose and will be exposed 
to a neutron flux of $6\times 10^{11}$ n/cm$^{2}$ in three years of 
running~\cite{MCirr}. For this reason, we have tested the variation 
of the light yield (LY) and the longitudinal response uniformity (LRU) of un-doped CsI
crystals irradiated with a dose and a neutron flux
exceeding the one expected in the experiment lifetime.

Concerning SiPMs, they have to withstand a fluency of $3\times10^{11}$~n/cm$^{2}$,
with neutrons at 1~MeV equivalent energy, 
and absorb a dose up to 20 krad in three years of 
running~\cite{MCirr}.

Irradiation tests with a ionization dose have been performed at the 
ENEA CALLIOPE facility~\cite{CALLIOPE}, where a $^{60}$Co source is 
used to produce $\gamma$'s with an energy of 1.25 MeV. The activity of the source during our tests was 
$0.35\times 10^{15}$ Bq, allowing to reach 5 Gy/h at about 
5 m distance.

The neutron irradiation tests have been performed at the ENEA 
FNG facility \cite{FNG}, where a nearly isotropic 14 MeV neutron flux is produced. 
The maximum neutron intensity is $0.5\times 10^{11}$~n/s, close to 
the target, with a uniform production and a dependence on the distance, 
$R$, as $1/R^2$. The desired neutron intensity is reached by 
positioning the crystal at the needed distance.

\section{Pure CsI radiation hardness measurements}
In both irradiation tests, the crystals parameters, LY and LRU, have been measured
 at different steps of the irradiation program. 
This characterization is done using a low intensity collimated $^{22}$Na source. 
The source is placed between the crystals and a small monitor system, 
constituted by a ($3\times 3\times 10$) mm$^3$ LYSO crystal, readout by a 
($3\times 3$) mm$^2$ MPPC. One of the two back-to-back 511 keV photons 
produced by the source is tagged by this monitor while the second 
photon is used to test the crystal, that is readout by means of a 2'' 
UV-extended 9813QB photomultiplier (PMT) from ET Enterprises. 
The LY has then been evaluated as:
\begin{equation}
\frac{Np.e.}{MeV} = \frac{\mu_{Q [pC]}}{G_{PMT}\cdot E_{\gamma} [MeV] \cdot q_{e^-} [pC]},
\end{equation}
where $Q$ is the charge obtained integrating and converting in pC the signal amplitude and $\mu_Q$ is the mean value provided by the gaussian fit applied, 
$G_{PMT}=3.8\times 10^6$ is the PMT gain, 
$q_{e^-}=1.6\times10^{-19}$~pC is the charge of the electron and  $E_\gamma = 511$~keV 
is the energy of the annihilation photon. 

A $(3\times 3\times 18)$ cm$^3$ CsI crystal from SICCAS has been 
irradiated at the CALLIOPE facility, by positioning it parallel 
to the incoming $\gamma$'s, so that the same dose was delivered 
along the crystal axis. The total dose 
absorbed by the crystal was $\sim 90$ krad (900 Gy, two days at 2.23 Gy/h and seven days at 4.86 Gy/h), that corresponds 
to about three times the total dose expected in the hottest region
of the calorimeter, i.e.\ the innermost ring around the beam axis,
in three years running \cite{MCirr}.

\begin{figure}[htbp]
        \centering
       
        \begin{minipage}[c]{.45\textwidth}
          \centering
          \includegraphics[width=\textwidth]{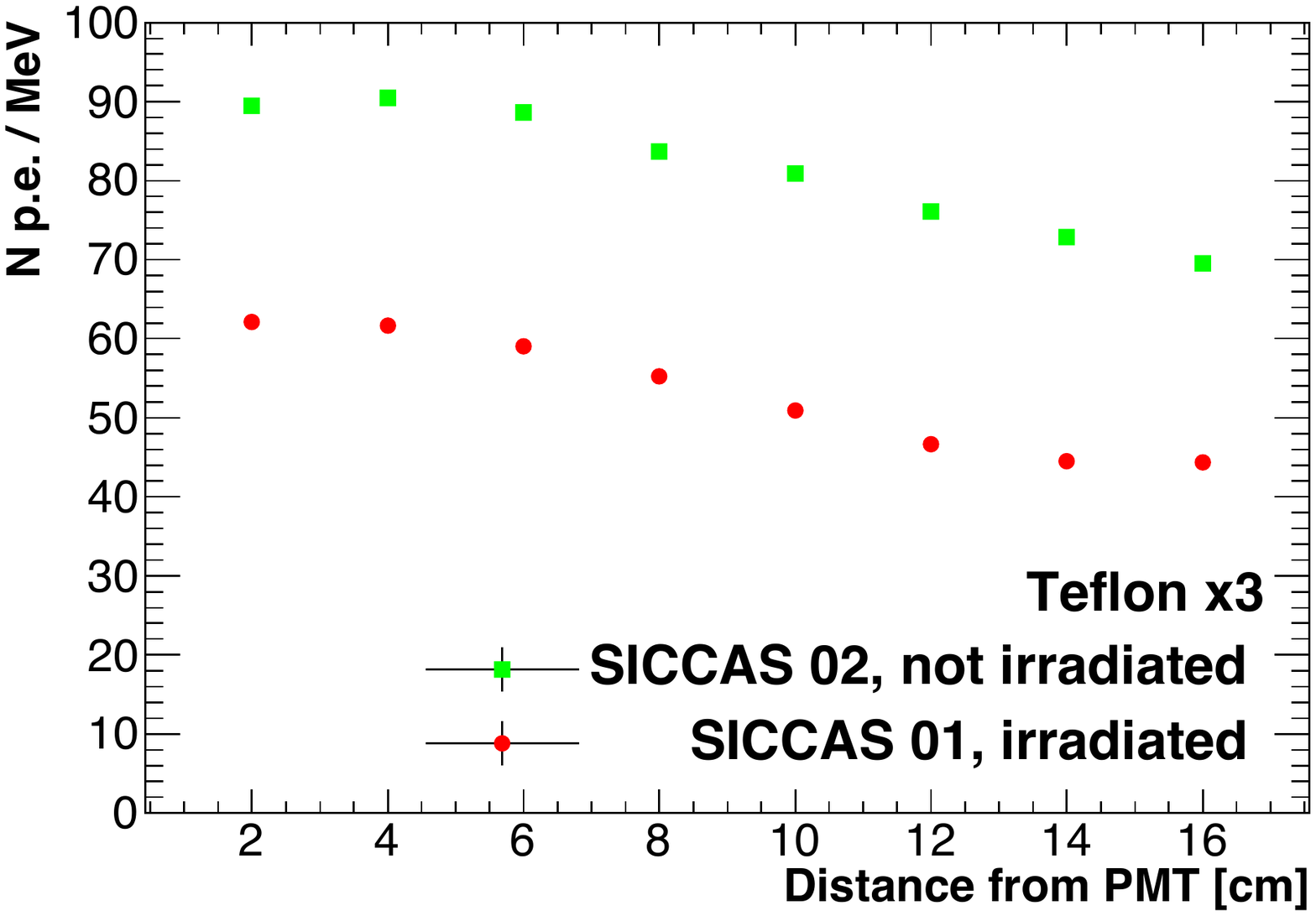}

        \end{minipage}%
        \hspace{9mm}%
        \begin{minipage}[c]{0.45\textwidth}
          \centering
          \includegraphics[width=\textwidth]{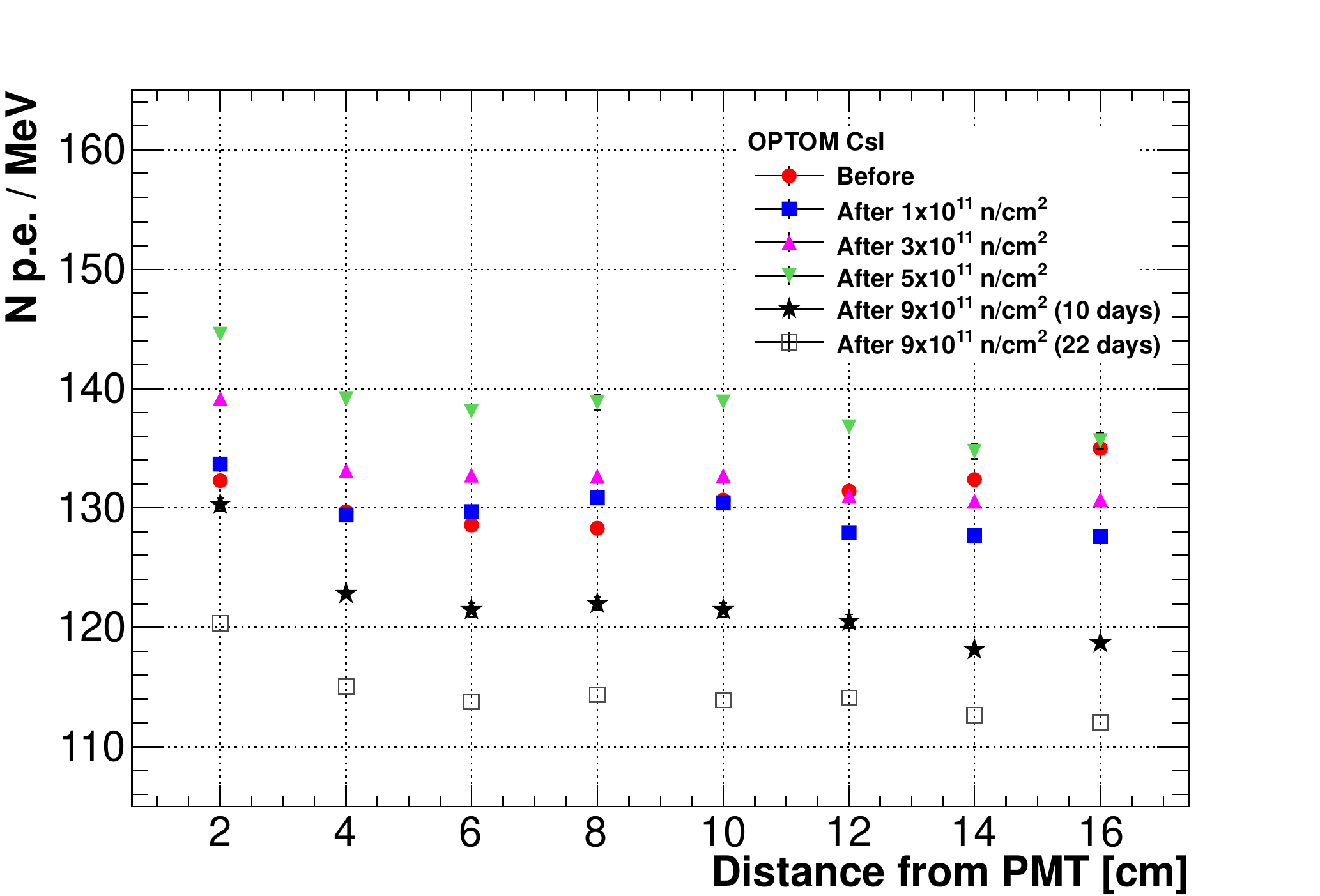}
         \end{minipage}
      
         \caption{Crystal LY as a function of the source distance from the PMT at different steps of the irradiation programs at CALLIOPE (left) and FNG (right).  \label{Fig:LYgamma}}
      \end{figure}

As shown in Figure~\ref{Fig:LYgamma} (left), a negligible LY reduction is 
observed for doses up to 20 krad, while for the last irradiation  
step the measured decrease is about 20\%. 
The slope for the LRU remains similar for the irradiated and not irradiated crystals.

Three $(3\times 3\times 20)$ cm$^3$  pure CsI crystals from different vendors have been 
tested with neutron dose at FNG: one from SICCAS (China), one from ISMA (Ukraine) and one from OPTO MATERIALS (Italy).
The total flux delivered, 
$9\times 10^{11}$ n/cm$^{2}$, which corresponds to about 1.5/3 times the 
maximum total flux expected for the first/second calorimeter disk 
in three years of running \cite{MCirr}. LY and LRU for Tyvek 
wrapped crystals and PMT optically coupled in air have been measured 
before each day of irradiation. The measurements have also been 
repeated 10 and 22 days after the end of the irradiation test.

In Figure~\ref{Fig:LYgamma} (right), an example of the LY of one crystal tested
is reported as a function of the source distance from the PMT. The LY increases
with the neutron flux because of fluorescence effects and activation.
Measurements performed several days after the irradiation test show
a decrease of the LY as related to the reduction of these
effects. Comparing the first to the last measurements, a decrease of 
10-20\% is visible for the undoped CsI crystal from OPTO MATERIALS, 
while no deterioration is present for the ISMA one. The SICCAS 
crystal has a completely different behavior along the crystal before 
and after irradiation, with a strong deterioration of the uniformity 
slope. Moreover, all crystals, except for SICCAS one, after 
irradiation show good LRU performance, with a total uniformity well 
below 10\%, and negligible deterioration. LRU for CsI from SICCAS 
changes from 5\% to 15\% after neutron irradiation.

\section{Silicon Photomultiplier radiation damage induced}
During the irradiation campaign different models of SiPM have been tested: 
two SiPM from Hamamatsu \cite{Hamamatsu} and a SiPM from FBK \cite{FBK}.
The (12 $\times$ 12)~mm$^2$ Hamamatsu SiPMs are made by the same array of
16  ($3 \times 3$) mm$^2$ cells, but they have different protection material: 
one SiPM is covered with a silicon protection layer (SPL) 
while the other one with a Micro-Film (MF).
The FBK SiPM is instead a monolithic (6 $\times$ 6)~mm$^2$ SiPM.

To control and monitor the Hamamatsu devices, a pulsed UV-led and two optical 
fibers have been used to illuminate the SiPM and a PMT positioned 2 m far away from the 
radiation source. We acquired the signal response to the led pulse of just one 
cell and the leakage current of another cell. The PMT response 
has been used as reference in order to obtain a precise measurement of the light input. 
Since the FBK SiPM is monolithic only its leakage current has been recorded.

\begin{figure}[h!]
        \centering
        \begin{minipage}[c]{.45\textwidth}
          \centering
           \epsfig{file=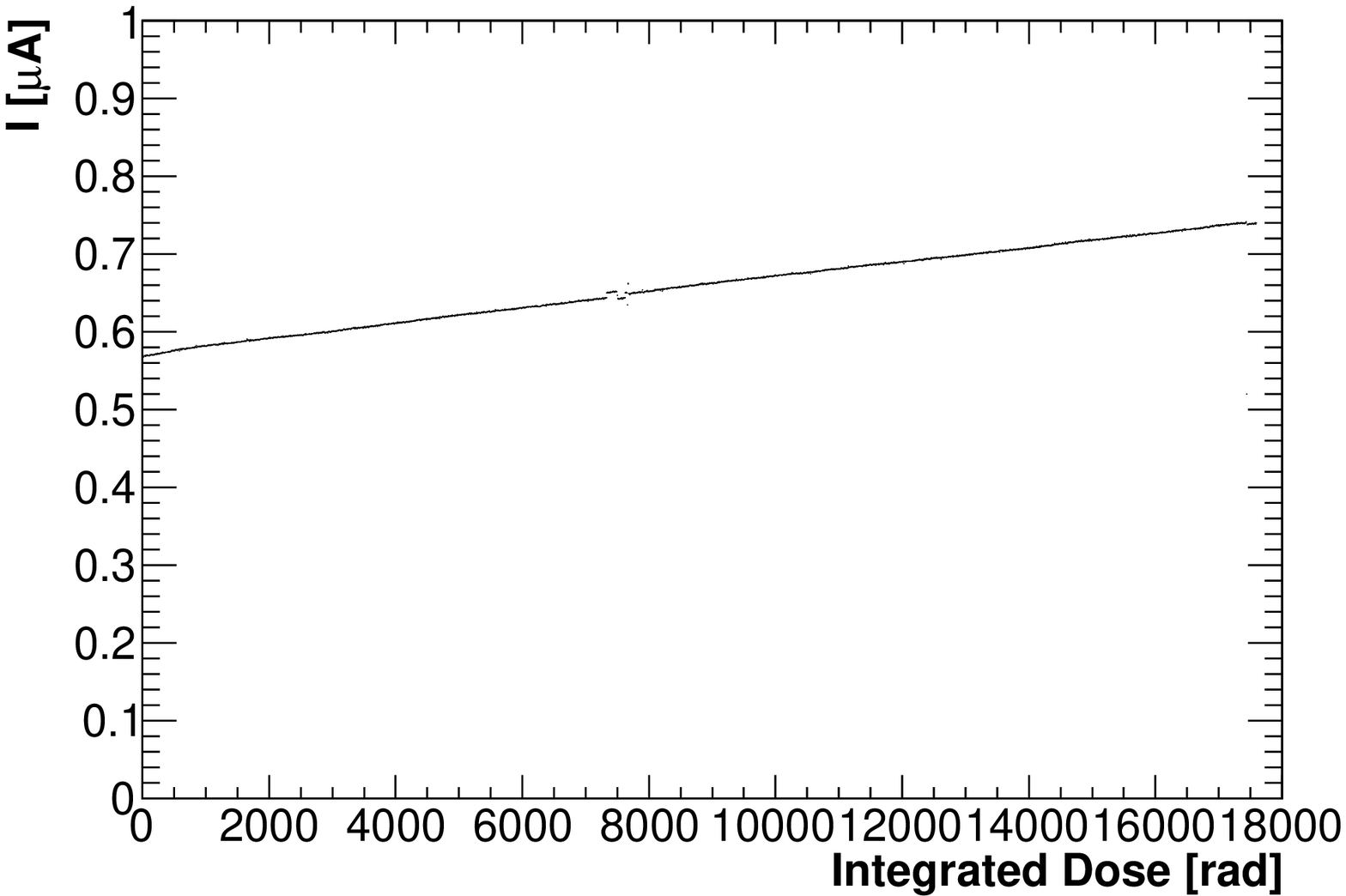,width=\textwidth}
        \end{minipage}%
        \hspace{9mm}%
        \begin{minipage}[c]{0.45\textwidth}
          \centering
          \includegraphics[width=\textwidth]{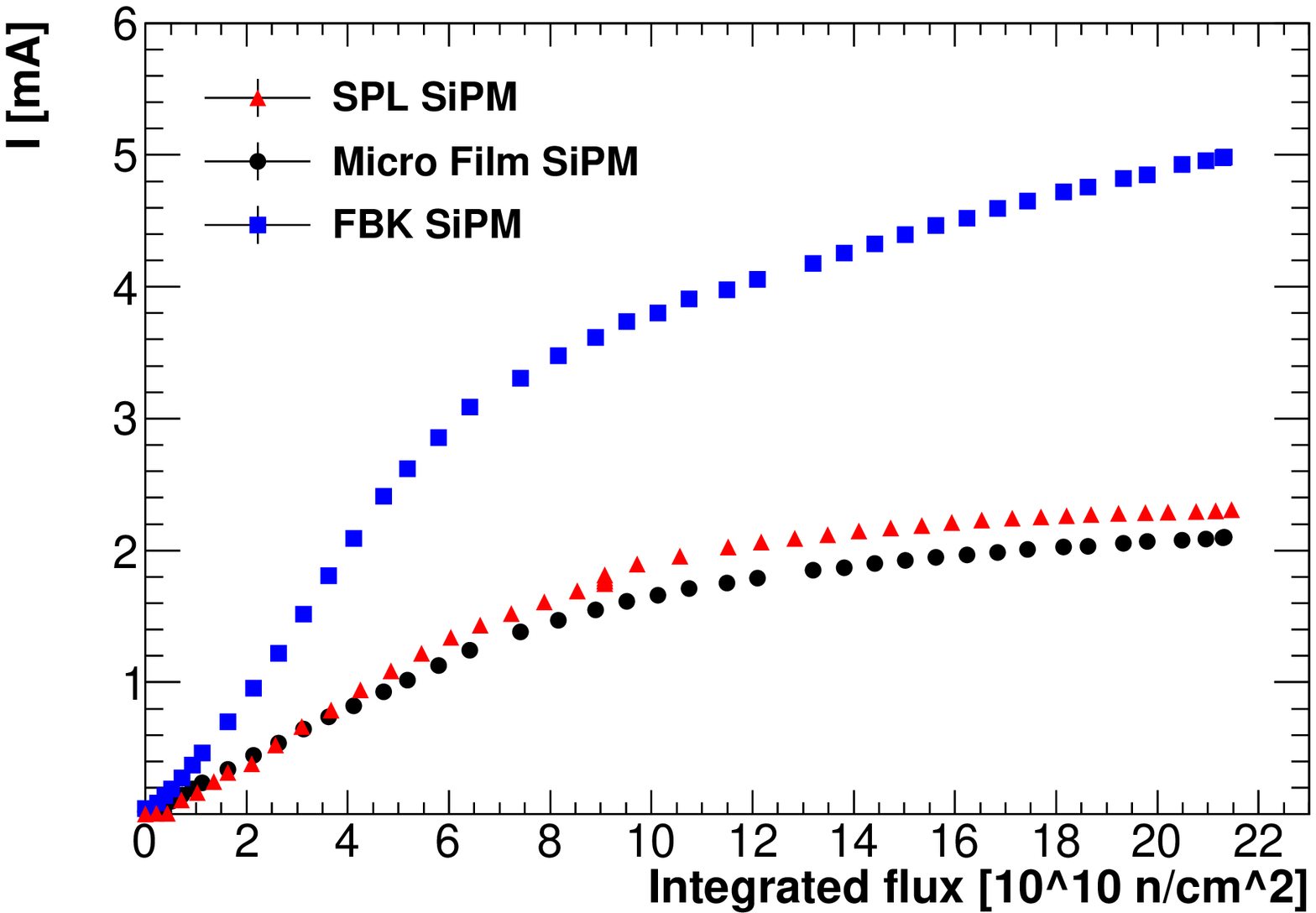}
         \end{minipage}
         \caption{Leakage current as a function of the integrated dose absorbed by the SPL SiPM at CALLIOPE (left) and as a function of the neutron integrated flux at FNG (right).  \label{Fig:I_flux}}
      \end{figure}

The three SiPMs were first irradiated with neutrons at the FNG facility. 
They were positioned 7~cm far away from the source and irradiated, for 
less than 4 hours, with 14~MeV neutrons up to $2.2 \times 10^{11}$~n/cm$^2$, 
which corresponds to a neutrons flux equal to 2.2 times that expected in the experimental lifetime.
We observed that the signal peak of SPL SiPM decreased from $\sim$~250~mV to $\sim$~30~mV, 
while the MF one decreased from $\sim~400$~mV to $\sim~50$~mV.
In Figure~\ref{Fig:I_flux} (right)  the leakage current of all the SiPMs tested as a function of the 
integrated flux is reported. To compare results from  SiPMs with different cell dimensions, the 
FBK current has been corrected by a factor 4 because of the greater active area.
A current increase is clearly visible in both SiPMs: the leakage current of MF SiPM increased from 
$\sim16~\mu$A to $\sim2$~mA, the one of SPL SiPM from $\sim100~\mu$A to $\sim2.2$~mA 
and of FBK one from  $\sim~21~\mu$A to $\sim~5$~mA. Even if the hall temperature was quite 
stable during irradiation the drop on the gain was mostly dominated by the temperature increment of the SiPM.

Later, another SPL SiPM has been irradiated at CALLIOPE, with a total dose 
$\sim$~20 krad in 3 days. The dose effect on SiPM performance is negligible
both in term of leakage current and signal amplitude. As shown in Figure.~\ref{Fig:I_flux} (left)
the leakage current increased from $\sim~0.6~\mu$A to $\sim~0.75~\mu$A  while the signal 
amplitude remained unchanged after the irradiation.

Other irradiation tests with SiPM kept at stable temperature are foreseen.

\section{Conclusions}
The determination of the LY and LRU changes for un-doped CsI crystals after irradiation with a large ionization dose and with neutron fluency provides an important benchmark for the Mu2e calorimeter, where a high radiation environment is foreseen. Our tests show that doses up to 200 Gy do not modify LY and LRU for a $(3 \times 3 \times 18)$ cm$^3$ unwrapped CsI crystals from SICCAS coupled in air to an UV-extended PMT. After a total dose of 900 Gy, a 20\% reduction in LY is observed instead. Crystals from many vendors have also been irradiated with a neutron flux of $9 \times 10^{11}$ n/cm$^2$, corresponding to about 2 times the total flux expected for the hottest calorimeter regions in three years of running. At the end of the irradiation test, an acceptable deterioration for the LY has been observed and the LRU is maintained well below 10\%.

We have also tested the radiation damage of the Hamamatsu SPL and Micro Film SiPMs, as well as FBK SiPMs, with neutrons and photons by measuring their change in response and leakage current. The total neutron flux ($2.2 \times 10^{11}$) causes a decreas of the signal peak (and gain) and a large increase of the leakage current.  A dose up to 200 Gy causes a negligible effect. Changes are still acceptable for the running conditions in the experiment when cooling down the SiPM to a running  temperature of $\sim$ 0 $^\circ$C. 

\ack
This work was supported by the EU Horizon 2020 Research and Innovation Programme under the Marie Sklodowska-Curie Grant Agreement No. 690835.

\end{document}